\documentclass[3p,times]{elsarticle}

\usepackage{ecrc}

\volume{00}

\firstpage{1}

\journalname{Nuclear Physics A}

\runauth{C. Gale}

\jid{nupha}

\usepackage{amssymb}
\usepackage{url}
\usepackage[figuresright]{rotating}

\begin{document}

\begin{frontmatter}



\dochead{}

\title{Electromagnetic radiation in heavy ion collisions: Progress and puzzles}

\author{Charles Gale}

\address{Department of Physics, McGill University\\ 3600 rue University, Montreal, QC, Canada H3A 2T8}

\begin{abstract}
We review the current state of photon and dilepton measurements at RHIC, emphasizing that of the theoretical work seeking to interpret them. We highlight the progress made recently in the modelling of relativistic nuclear collisions, and explore the effect on electromagnetic observables. Some outstanding puzzles are presented.  
\end{abstract}


\end{frontmatter}

% \linenumbers

\section{Introduction}
\label{Intro}
The behaviour of strongly-interacting matter is known to be dictated by QCD, and colliding large nuclei  at relativistic energies has been convincingly shown to be an efficient means of studying QCD under extreme conditions. A vibrant experimental program is currently under way at RHIC (the Relativistic Heavy Ion Collider, at Brookhaven National Laboratory), and at the LHC (the Large Hadron Collider, in CERN). In the analyses of experiments performed at those facilities,  measurements of the produced hadronic particles have revealed many fascinating features of the relativistic many-body systems.  Whereas  energetic jets  contain  information on the nature of the QCD plasma  at early times \cite{Tywoniuk,Renk}, the softer hadrons have shown an impressive final-state collectivity that has become a cornerstone of the  qualitative success of hydrodynamics at RHIC and LHC energies \cite{Schenke:2011qd,Luzum:2011mm}. Electromagnetic observables  complement well the hadronic measurements. Owing to the relative smallness of the fine-structure constant, $\alpha$, real and virtual photons emerge unscathed from their production site: They probe the entire space-time trajectory of the nuclear collision, from hot QCD plasma to the cooler hadronic phase. The price to pay for such a rich probe is a modest emission rate and a small signal-to-background ratio, in what concerns thermal photons, real and virtual. We review here some recent progress in the calculation of thermal photon yield and flow, and highlight some of the puzzles that have emerged in the analysis of electromagnetic radiation in relativistic heavy ion collisions. 
\section{Electromagnetic emission rates}

Throughout the history  of a single nuclear collision event, several sources of photons will come into play. The current formulation of relativistic hydrodynamics assumes that thermalization sets in after a proper time $\tau_0$. The value of this parameter is currently not calculable with certainty: it is left as a variable. Consequently, no reliable estimates exist for pre-equilibrium photons, other than those being generated in the very first instants of the collisions, calculable with perturbative QCD (pQCD), and measured in $pp$ collisions.  These ``prompt photons'' constitute an irreducible background to the photons being generated through some influence of a thermal medium. This statement, however, should be clarified: pQCD photons have a jet fragmentation component which may be sensitive to quenching effects from the medium \cite{Turbide:2007mi}. The QCD jets interacting with the medium then convert into photons \cite{Fries:2002kt,Turbide:2005fk}. 
At $\tau > \tau_0$, the thermal components of the medium may also interact to generate photons. There, the appropriate respective emission rates, $R$, for real and virtual photons can be expressed using finite-temperature field theory \cite{GaleKapusta:2006} as 
\begin{eqnarray}
\omega \frac{d^3R}{d^3 {\bf k}} &=& - \frac{g^{\mu \nu}}{(2 \pi)^3} \frac{1}{e^{\beta \omega } - 1}\, {\rm Im}\, \Pi^{\rm R}_{\mu \nu} (\omega, {\bf k})\ , \nonumber \\
E_+ E_- \frac{d^6R}{d^3 {\bf p}_+ d^3 {\bf p}_-} &=& \frac{2 e^2}{(2 \pi)^6} \frac{1}{k^4}\left[ p_+^\mu p_-^\nu + p_+^\nu p_-^\mu - g^{\mu \nu} p_+ \cdot p_- \right] {\rm Im}\, \Pi_{\mu \nu}^{\rm R} (\omega, {\bf k}) \, \frac{1}{e^{\beta \omega} - 1}
\label{dileprate}
\end{eqnarray}
In the above, lepton masses have been neglected, ${\rm Im \, \Pi^{\rm R}}$ is the retarded photon self-energy, $\omega$ and ${\bf k}$ are components of the photon four-momentum $k$, $E_{\pm}$ and ${\bf p}_{\pm}$ are components of the lepton/antilepton four-momenta, and $\beta = 1/T$. Those expressions receive correction of ${\cal O} (\alpha^2)$ in the electromagnetic interaction, but are exact in the strong interaction. 

The finite-temperature photon emission rates in the quark-gluon plasma (QGP)  phase have been worked out to order $\alpha_s$, the leading order in the strong interaction, a little more than a decade ago \cite{Arnold:2001ba,Arnold:2001ms,Arnold:2002ja}. One may question the robustness of those rates against NLO corrections, especially in light of closely related investigations of the behaviour of higher-order correlation functions  \cite{CaronHuot:2008ni,CaronHuot:2009ns}. This has been done recently, and first results have been reported \cite{NLOphotons}. It is shown there, that the phenomenological consequences of the NLO photon-emission rates are but a modest $\sim$20\% enhancement in the region of momentum  $| {\bf k} |/T \approx 10$. It is inappropriate here to discuss the technical intricacies of this calculation, but suffice it to say that it therefore appears that  conclusions reached through  previous analyses relying on LO QGP rates need not be amended. From the theoretical point of view, however, this work represents definite progress. 

\begin{figure}[hb]
  \begin{center}
    \includegraphics[width=7cm]{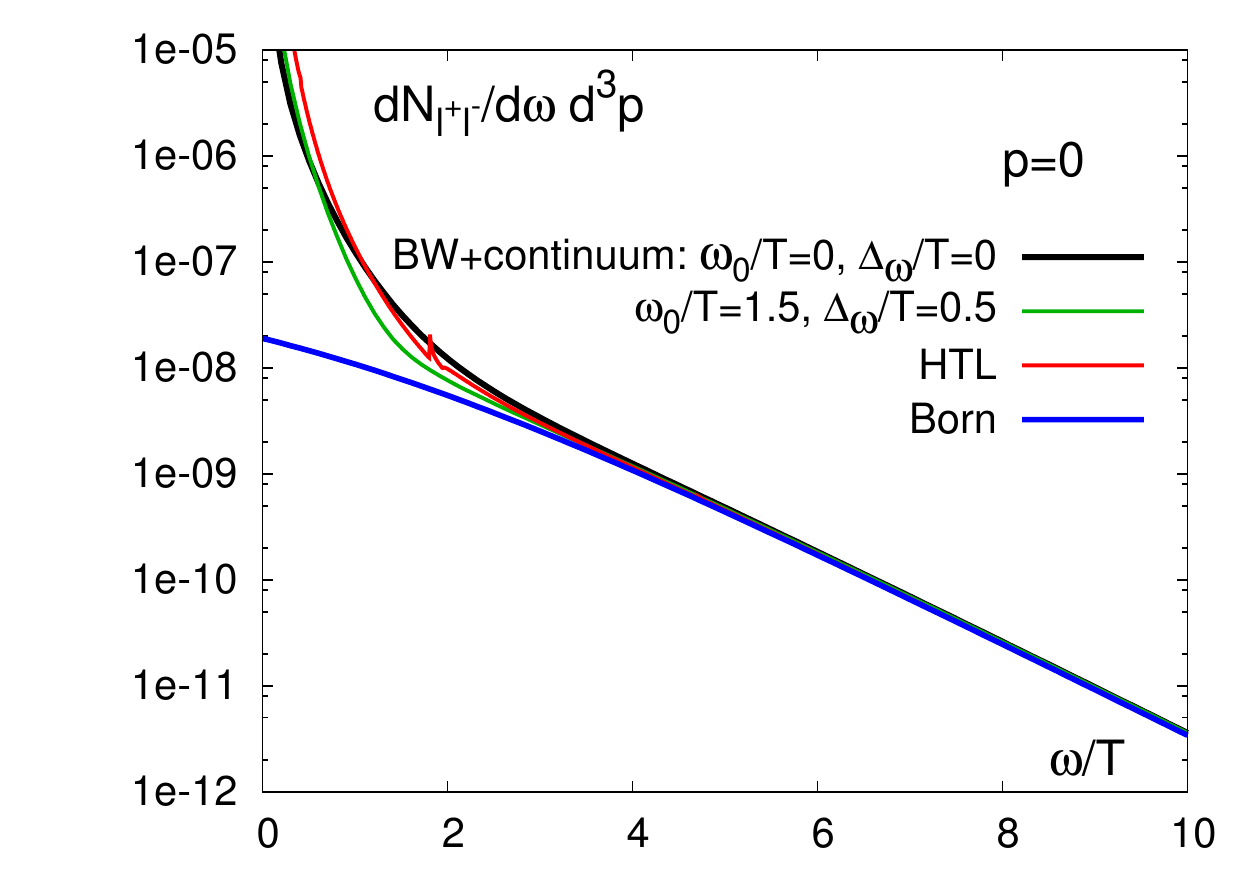}
    \caption{The rate for the production of back-to-back lepton pairs, calculated in quenched lattice QCD at $T \simeq 1.45\, T_c$, for two flavours of quarks.  The top two curves of the legend represent two different ways to extract the lattice data. Also shown are calculation done with the Hard Thermal Loop resummation \cite{Braaten:1990wp}, and with keeping only the Born term. This  plot is from Ref. \cite{Ding:2010ga}.}
    \label{lattdilep}
  \end{center}
  \vspace*{-.5cm}
\end{figure}
In what concerns dileptons, another recent element of  progress has been the nonperturbative evaluation of the vector current-current correlation function in lattice QCD at finite temperature.  The connection with the dilepton emission rate is obtained by straightforward manipulations of Eq. (\ref{dileprate}): 
\begin{eqnarray}
\frac{d^4 R}{d^4 k} = \frac{\alpha}{ 12 \pi^3} \frac{1}{M^2} f^{(n) \mu}_\mu (\omega, {\bf k}) \frac{1}{(e^{\beta \omega} - 1)}
\end{eqnarray}
where $f^{(n)}$ is the ``normal'' \cite{GaleKapusta:2006} spectral representation of the current-current correlator, and $M^2 = (p_+ + p_-)^2$.
Such extrapolations from lattice calculations however require solving an inversion problem that is mathematically ill-defined: The lattice calculates Euclidian correlation functions, $f ( \tau)$, formulated in imaginary time, $\tau$. The connection with the physical Minkowski correlation function $f (\omega)$, is
\begin{eqnarray}
f (\tau)) = \int_0^\infty d \omega f (\omega) \frac{\cosh \left( \beta \omega /2 - \tau \omega \right)}{\sinh \left( \beta \omega/2\right)}
\end{eqnarray}
Progress in inverting this (possibly noisy) integral has been obtained in recent years using  statistical approaches like the Maximum Entropy Method \cite{Asakawa:2000tr}. Some some recent work has involved using a physically-motivated ansatz for the lattice spectral function \cite{Ding:2010ga}. The dilepton production rate obtained in this fashion is shown in Figure \ref{lattdilep}. 
The lattice-extracted dilepton production rate merges with the Born rate at $\omega/T \sim 4$ and rises well over that, at low energies. Note that the low-$M$ HTL rate grows differently as the invariant mass is lowered: A known behaviour \cite{Karsch:2000gi,Moore:2006qn}. More results of this nature, but for a range of temperatures closer to the transition temperature and at finite three-momentum, will eventually open the door to a precise non-perturbative characterization of the dilepton emission rates in the QGP domain. 
On the hadronic side, photon and dilepton rates have mostly relied on using effective hadronic Lagrangians, supported by the existing data on hadronic radiative decays, nuclear photoabsorption, and electron-positron annihilation \cite{Rapp:1999ej,Rapp:1999qu,Turbide:2003si,Gale:2009gc}. The different hadronic medium electromagnetic emissivities are summarized in the cited literature reviews: Current approaches do differ in detail, but agree on the importance of broadened spectral densities driven by in-medium interaction.

Thus, there has been advances in the calculation of photon and dilepton emission rates. In addition to those, the field of relativistic heavy-on collisions has witnessed impressive leaps forward in the space-time modelling of the colliding system. As the electromagnetic rates need to be integrated to produce yields in order to be compared with experimental measurement, the space-time evolution needs to be considered on an equal footing. 

\section{Space-time modelling}

As already mentioned,  the success of relativistic hydrodynamics in the theoretical interpretation of relativistic nuclear collisions at RHIC and at the LHC is striking.   In addition to momentum distributions, particle spectra may be characterized by their dependence on the azimuthal angle $\phi$. This in turn will determine the momentum anisotropy, which can be quantified through a Fourier expansion \cite{Voloshin:2008dg}. The coefficients of this expansion, the flow coefficients, and the associated reaction plane are defined event-by-event:
\begin{eqnarray}
v_n = \langle \cos \left( n \left( \phi - \Psi_n \right) \right) \rangle\, , \ \ \ \ \   \ \ \ \ \Psi_n = \frac{1}{n} \arctan \frac{\langle \sin \left(n \phi \right) \rangle}{\langle \cos \left( n \phi \right) \rangle}
\end{eqnarray}
\begin{figure}[ht]
  \begin{center}
    \includegraphics[width=9.5cm]{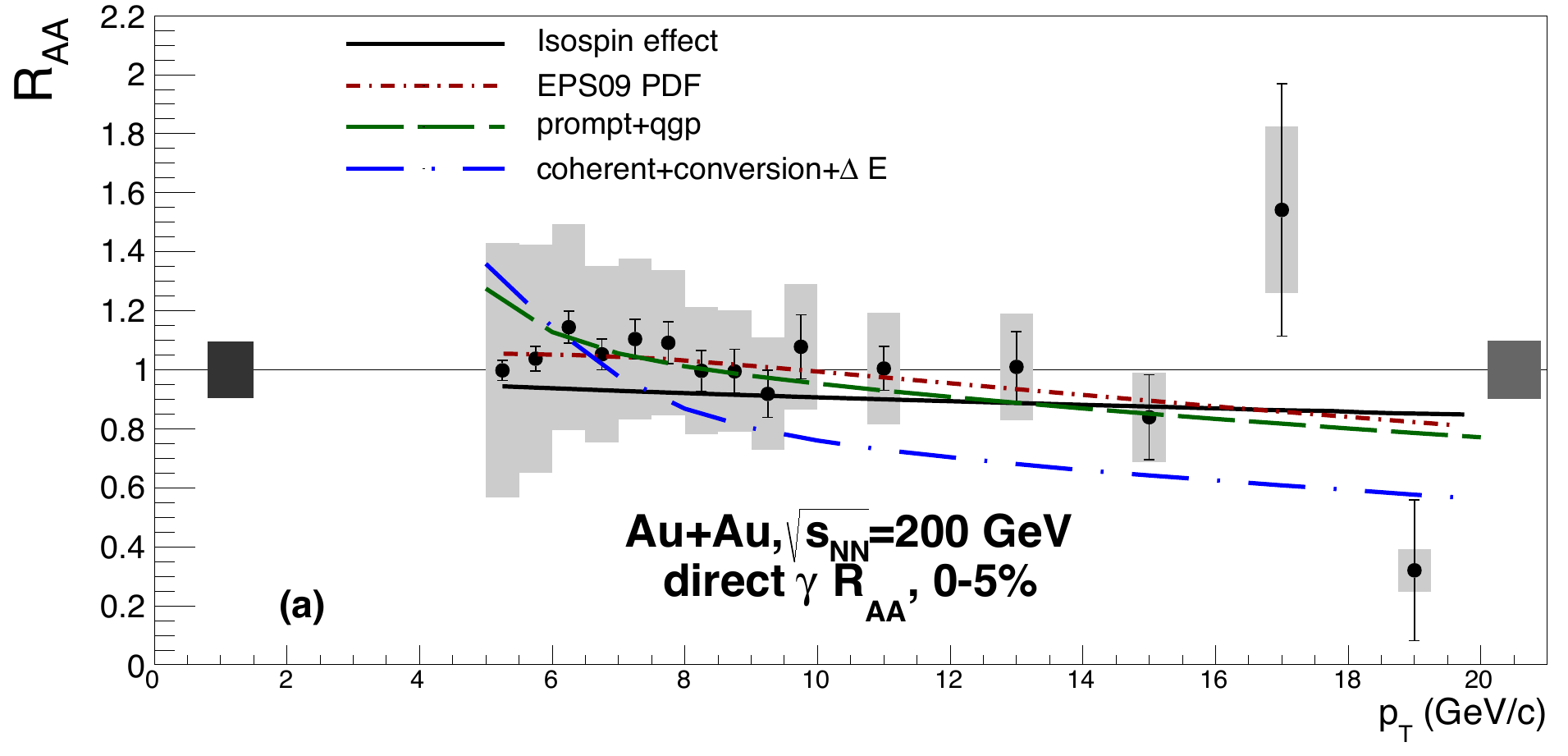} 
    \caption{The photon $R_{AA}$ is shown, as a function of the photon momentum, measured in Au + Au collisions at RHIC. The curves shown are results from different theoretical approaches. Those are discussed in the text and in the references.  The  plot is from Ref. \cite{Afanasiev:2012dg}.}
    \label{photRAA}
  \end{center}
  \vspace*{-.5cm}
\end{figure}
The recent progress in this area over the last few years has been such that the  possibility of extracting transport coefficients from relativistic heavy ion data is a real possibility. More specifically, a small but finite value of the shear viscosity to entropy density density ratio ($\eta/s$) appears consistent with a global fit to RHIC and LHC data sets, including measurements of higher flow harmonics \cite{schenke:2011bn}. 

Given the electromagnetic sources and emissivities discussed previously, what are the yields predicted by the modern hydrodynamical approaches? In what concerns real photons, a recent update \cite{Sahlmueller} of data analysis by the PHENIX collaboration  is shown in Figure \ref{photRAA}. What is plotted there is photon $R_{AA}$ at RHIC: the ratio of the measured yields in Au + Au collisions, divided by the production cross section in p + p collisions multiplied by the average nuclear thickness function for the appropriate centrality. The theoretical curves of Figure \ref{photRAA} relate to calculations that included effects that stem from cold nuclear matter (isospin and (anti)shadowing), effects of  photon generation by QGP and suppression by jet energy loss, and coherent effects such as LPM \cite{Afanasiev:2012dg}. As shown in Figure \ref{photRAA}, the data currently can't distinguish between an approach with no QGP \cite{Arleo:2011gc} and one which relies on 2+1D hydrodynamics where the photon-enhancing channels are almost canceled by the energy-loss effects \cite{Turbide:2007mi,Sahlmueller:2012fb}. Finally, an analysis of the low-momentum photon data, extracted from continuing dilepton measurements to the light-cone, has produced an effective temperature of the photon signal that exceeded pQCD expectations    \cite{Adare:2009qk}. For central collisions: $T = 221 \pm 19^{\rm stat} \pm 19^{\rm syst}$ MeV. Even if the spread in initial temperatures in the theoretical models that seek to explain the same data is considerable, $ 300 < T_{\rm init} {\rm (MeV)} < 600$, a consistent feature is that those values all exceed the typical transition temperature predicted by lattice QCD, as does the empirical value extracted from data. A critical review of the theoretical calculations that led to this wide palette of initial temperatures is due, and should reduce this scatter considerably.

Dilepton spectra have been available at RHIC for some years. The first measurements there were done by the PHENIX collaboration. A striking feature of these  is a large enhancement over background in the low invariant mass region: $0 < M ({\rm{GeV}}/c^2) < 0.7$, in Au + Au collisions \cite{Adare:2009qk}. The enhancement factor over the hadronic cocktail expected from p + p measurements has been measured to be $4.7 \pm 0.4^{\rm stat} \pm 1.5^{\rm syst} \pm 0.9^{\rm model}$ in minimum bias events, and this enhancement is concentrated at low $p_T (p_T < $1 GeV).  Since these pioneering data were recorded, it is fair to write that they have constituted a challenge for theory. This mismatch between theory and experiment persists to this day - models underpredict the data typically by a factor $\sim 3$ at $M \sim 400$ MeV -   and currently represents a puzzle.

\begin{figure}[ht]
  \begin{center}
  \vspace*{-1.5cm}
  \begin{minipage}[h]{6.9cm}
\includegraphics[width=6.9cm]{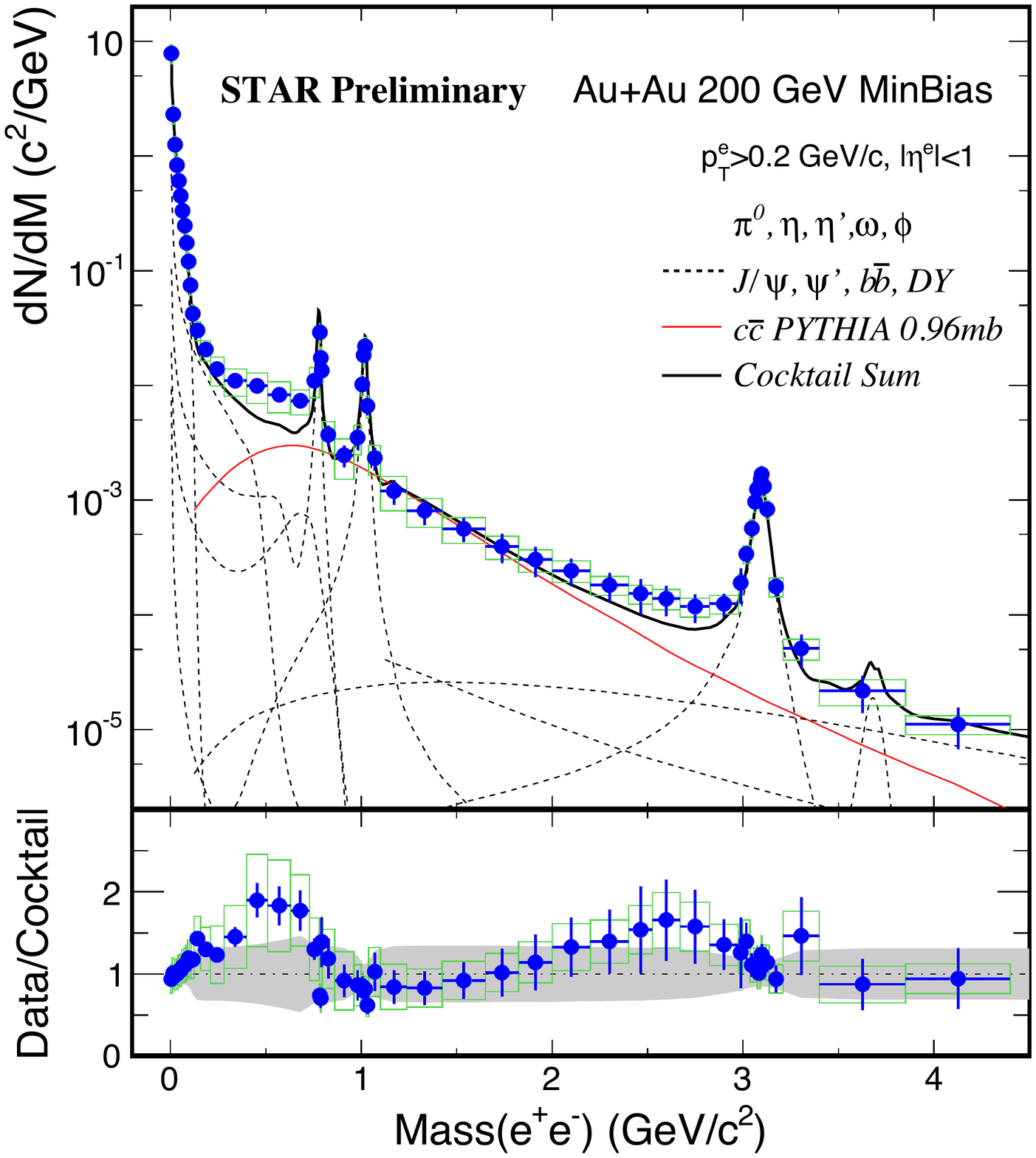}
\end{minipage}
\begin{minipage}[h]{7cm}
\vspace*{0.15cm}
\includegraphics[width=7cm]{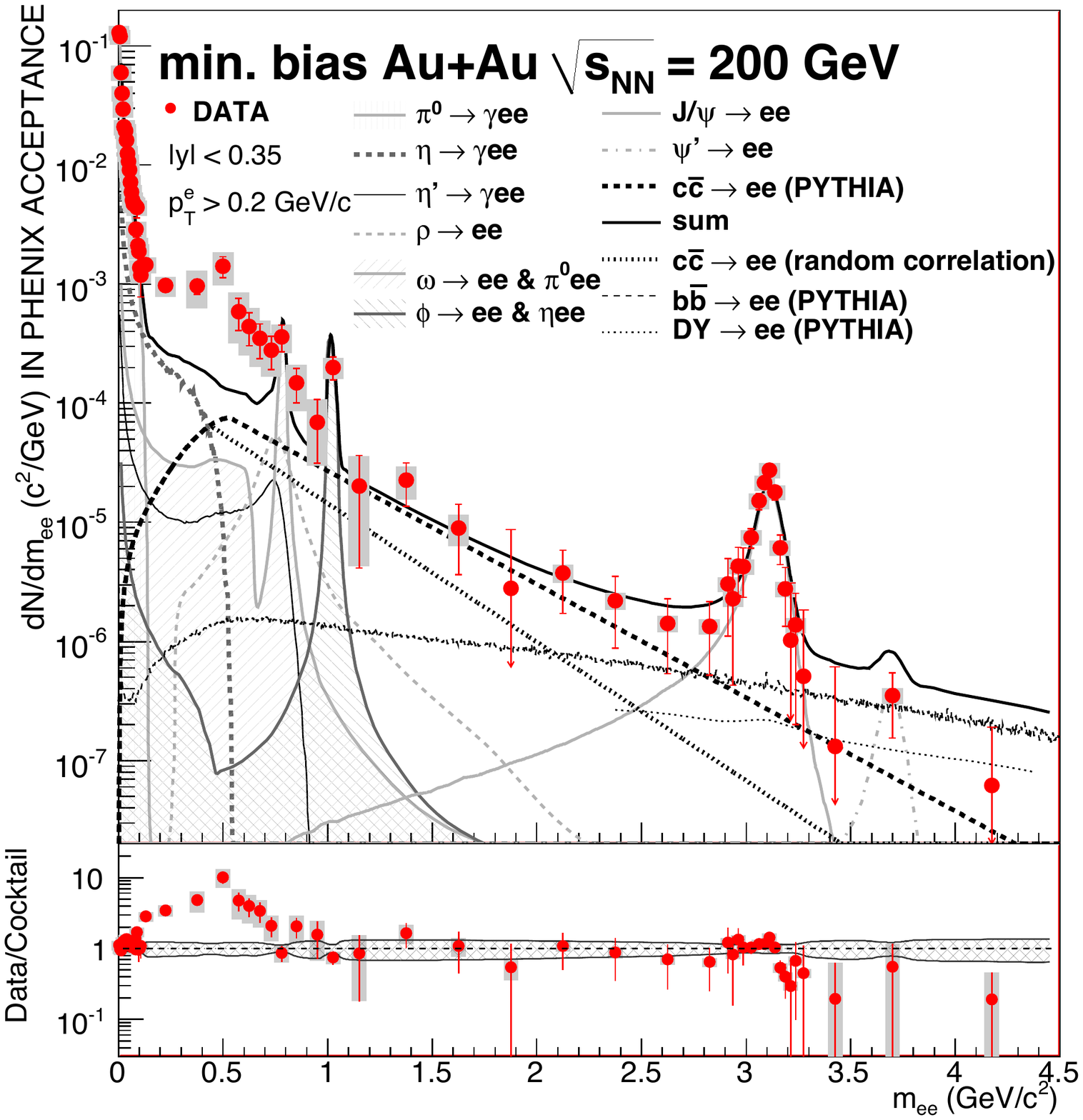} 
\end{minipage}
    \vspace*{-1.3cm}
    \caption{The yield of low mass dilepton as measured at RHIC by STAR (left panel) and by PHENIX (right panel).  Also shown is the data to cocktail ratio, for the two respective experiments. The left panel is from Ref. \cite{Lijuan}, and the right panel from Ref. \cite{Afanasiev:2012dg}.}
    \label{STARdilep}
  \end{center}
      \vspace*{-.5cm}
\end{figure}
Another element of experimental progress in the field has been the entrance of the STAR collaboration on the dilepton scene, owing to the recent installation of a Time-of-Flight detector. Both the TOF and the Time Projection Chamber (TPC) enable electron identification from low to intermediate $p_T$. The recent STAR dilepton measurements are shown in Figure \ref{STARdilep}. This collaboration also reports an enhancement (over the expected hadronic cocktail) in the low dilepton invariant mass region. However, this enhancement is smaller than the corresponding PHENIX figure: the STAR factor is $1.53 \pm 0.07^{\rm stat} \pm 0.41^{\rm syst}$. Results are still preliminary, but it appears unlikely that this difference in the enhancement factor observed by the two different collaborations is due to acceptance effects \cite{Drees}; it is thus currently also  a puzzle. Much is expected  in the low invariant mass region from analyses already made with PHENIX's Hadron Blind Detector, designed to reduce background by suppressing Dalitz pairs and photon conversions \cite{Anderson:2011jw}.

\subsection{Viscous effects}
\begin{figure}[h]
  \begin{center}
    \includegraphics[width=7cm]{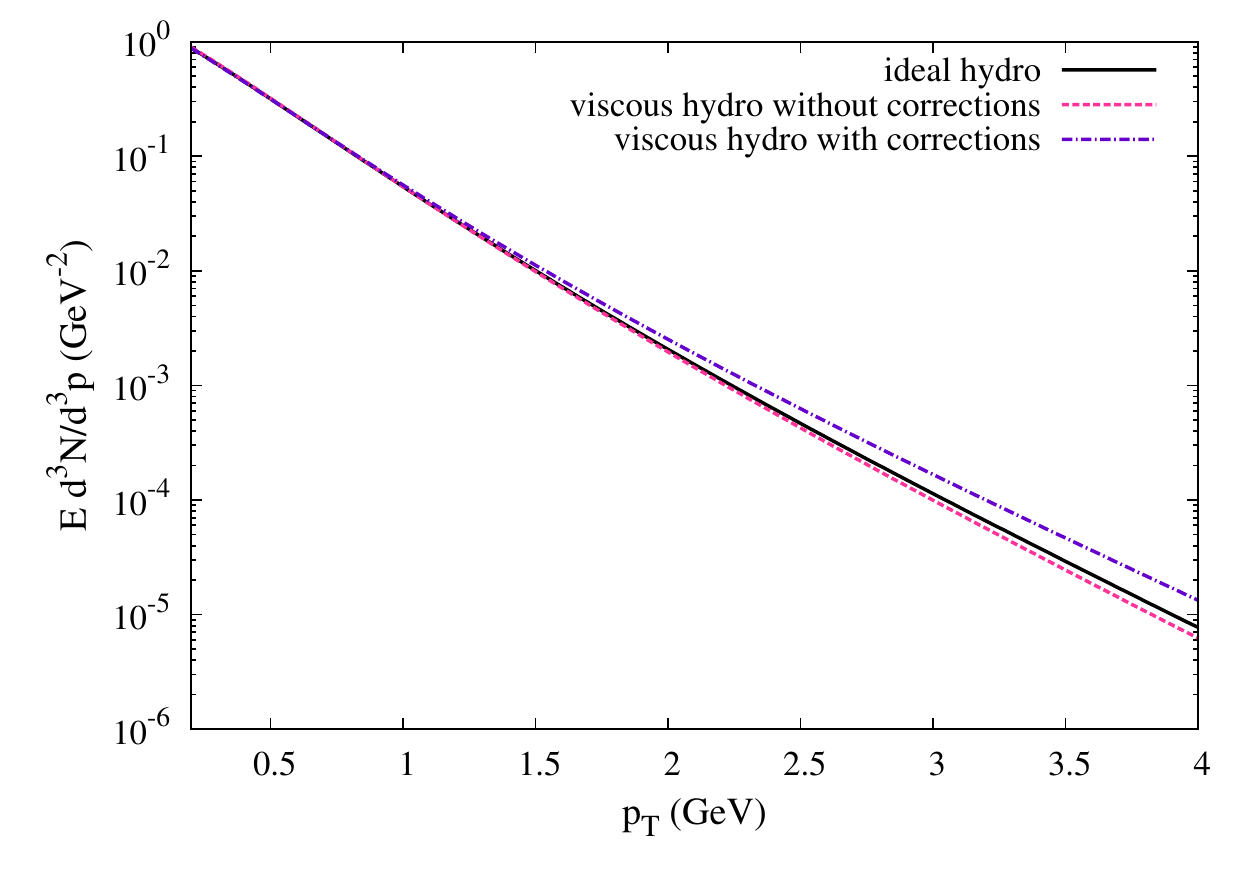}
        \includegraphics[width=7cm]{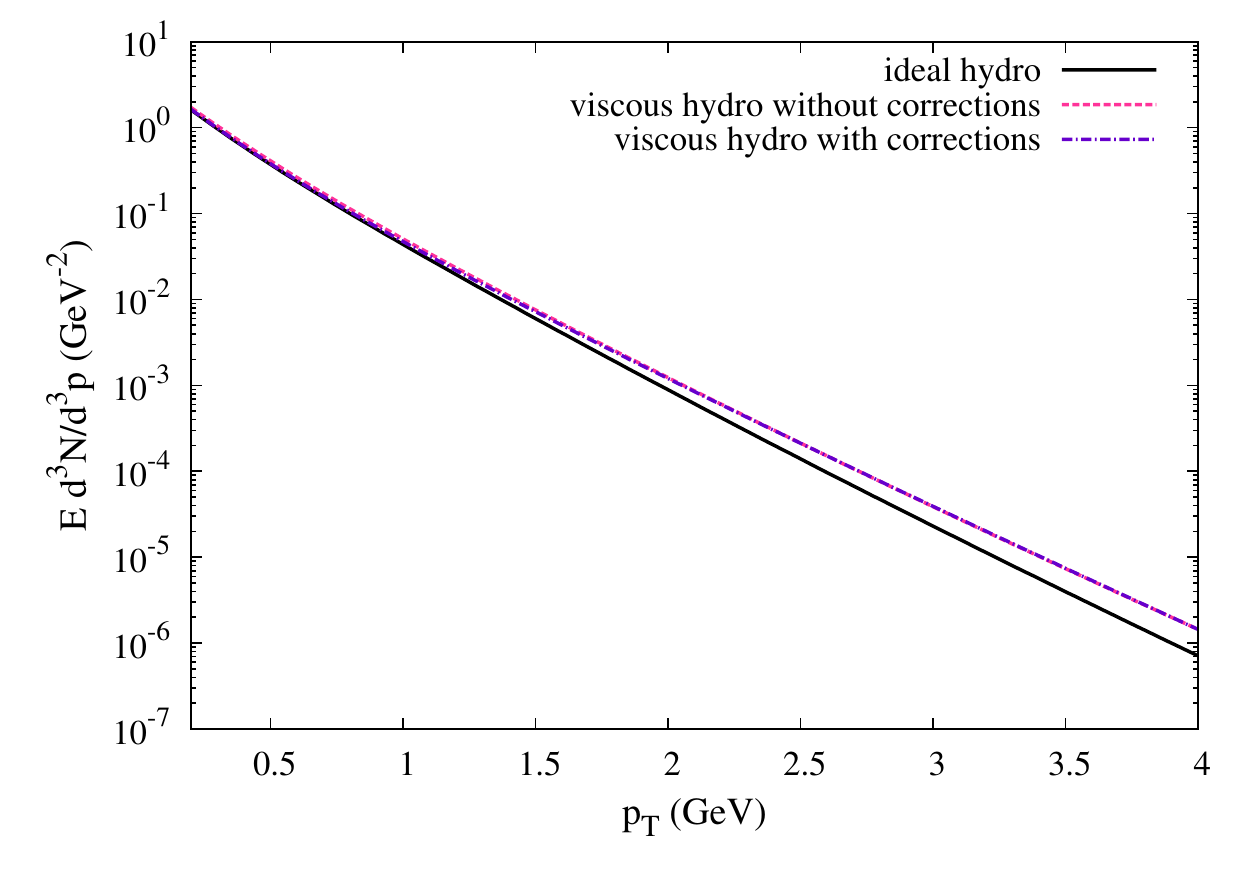}
    \caption{The photon yields, after integrating the rates through the hydrodynamic evolution. The full curves are results obtained with inviscid hydrodynamics, the dotted lines represent yields obtained with viscous hydro, but using uncorrected rates, and the dashed-dotted curves are yields with the viscous hydro and with the appropriately modified emission rates. A value of $\eta/s = 1/4\pi$ has been used. Left panel: Photons produced from the parton phase only. Right panel: Photons from the hadronic gas only.  This  plot is from Ref. \cite{Dion:2011pp}.}
    \label{visc_phot}
  \end{center}
  \vspace*{-.5cm}
\end{figure}

The last few years have seen the advent of fully three-dimensional hydrodynamics simulations, with viscous corrections up to second order in flow velocity gradients: see the previously cited reviews (\cite{Schenke:2011qd,Luzum:2011mm}), and references therein. We will only discuss here the effect of shear viscosity; How will it affect the generation of electromagnetic radiation? The introduction of a finite value of the shear viscosity coefficient will modify the microscopic distribution functions. In a multispecies ensemble, a popular ansatz for modifying the equilibrium distribution function of species $i$ is 
\begin{eqnarray}
f_{0 i} \to f_{0 i} + \delta f_i \, , \ \ \delta f_i = f_{0 i} ( 1 \pm f_{0 i} ) p^\alpha p^\beta \pi_{\alpha \beta} \frac{1}{2 \left(\epsilon + P \right) T^2}\ ,
\label{visc}
\end{eqnarray}
where $\pi_{\alpha \beta}$ is the shear correction to the stress-energy tensor, $\epsilon$ is the energy density, and $P$ is the pressure. This ansatz is however not unique \cite{Dusling:2009df}. In the QGP phase, the complete leading-order rates in $\alpha_s$ have yet to be re-evaluated, but the viscous corrections to the Compton and annihilation contributions have been done \cite{Schenke:2006yp,Dusling:2009bc,Chaudhuri:2011up,Dion:2011pp}. In the hadronic (confined) sector, the evaluation of viscous corrections to photon emission rates requires a recalculation of the many different channels that went into the compilation of Ref. \cite{Turbide:2003si}, for example, using Eq. (\ref{visc}). The viscous corrections to the photon yields thus manifest themselves through a modified hydrodynamic evolution, and also by altering the microscopic rate equations themselves \cite{Dion:2011pp}. The respective and combined effects are shown in Figure \ref{visc_phot}, for conditions corresponding to Au + Au collisions at $\sqrt{s} = 200$ GeV, and a $0 - 10 \%$ centrality class. 
For the partonic phase photons, using the viscous hydro (without changing the rates) reduces the yield at high $p_T$: the viscous hydro requires a lower initial temperature (than ideal hydro)  to be compatible with hadronic data, as entropy is generated throughout the evolution. For the hadronic gas photons, the main effect of viscosity is to produce a flatter temperature-against-time evolution \cite{Dion:2011pp}, and the higher late-stage temperatures will then produce more photons. From Eq. (\ref{visc}), the viscous corrections to the rates are proportional to momentum and to the shear pressure tensor. The tensor components are maximal at small times and high temperatures, driven by the large initial gradients, and are much smaller at later times \cite{Dion:2011pp}. This supports the net viscous yields in the hadronic gas phase not being noticeably affected by the viscous corrections to the microscopic processes. Conversely, the partonic  photons receive a correction which grows with momentum, making the spectra harder \cite{Dusling:2009bc,Bhatt:2010cy,Chaudhuri:2011up}.  However, this hardening of the spectra is modest, building up  to a factor of $\approx 2$ at $p_T \sim 4$ GeV. The pQCD photons will take over in this momentum region, making the extraction of a viscosity from photon spectra a challenging task. 

\begin{figure}[ht]
  \begin{center}
    \includegraphics[width=7cm]{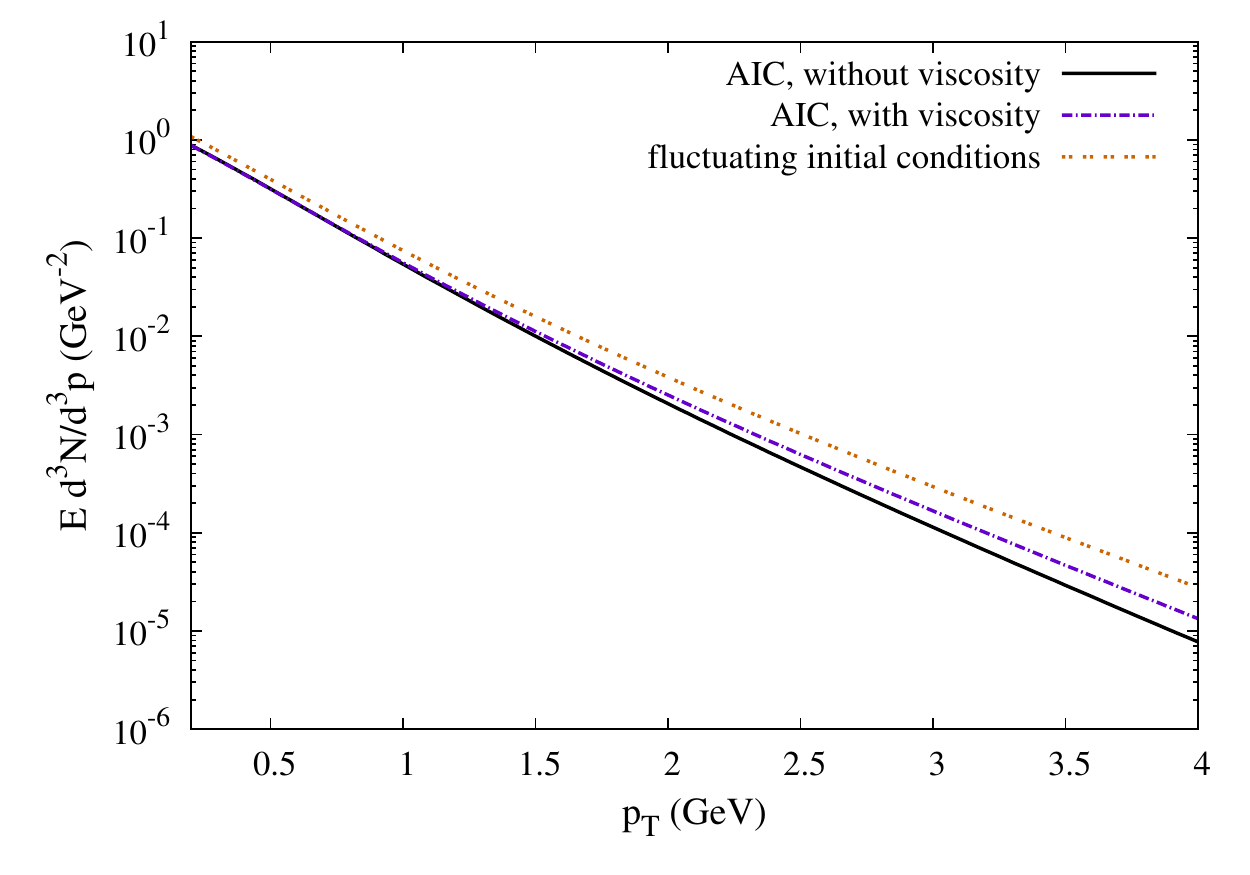}
        \includegraphics[width=7cm]{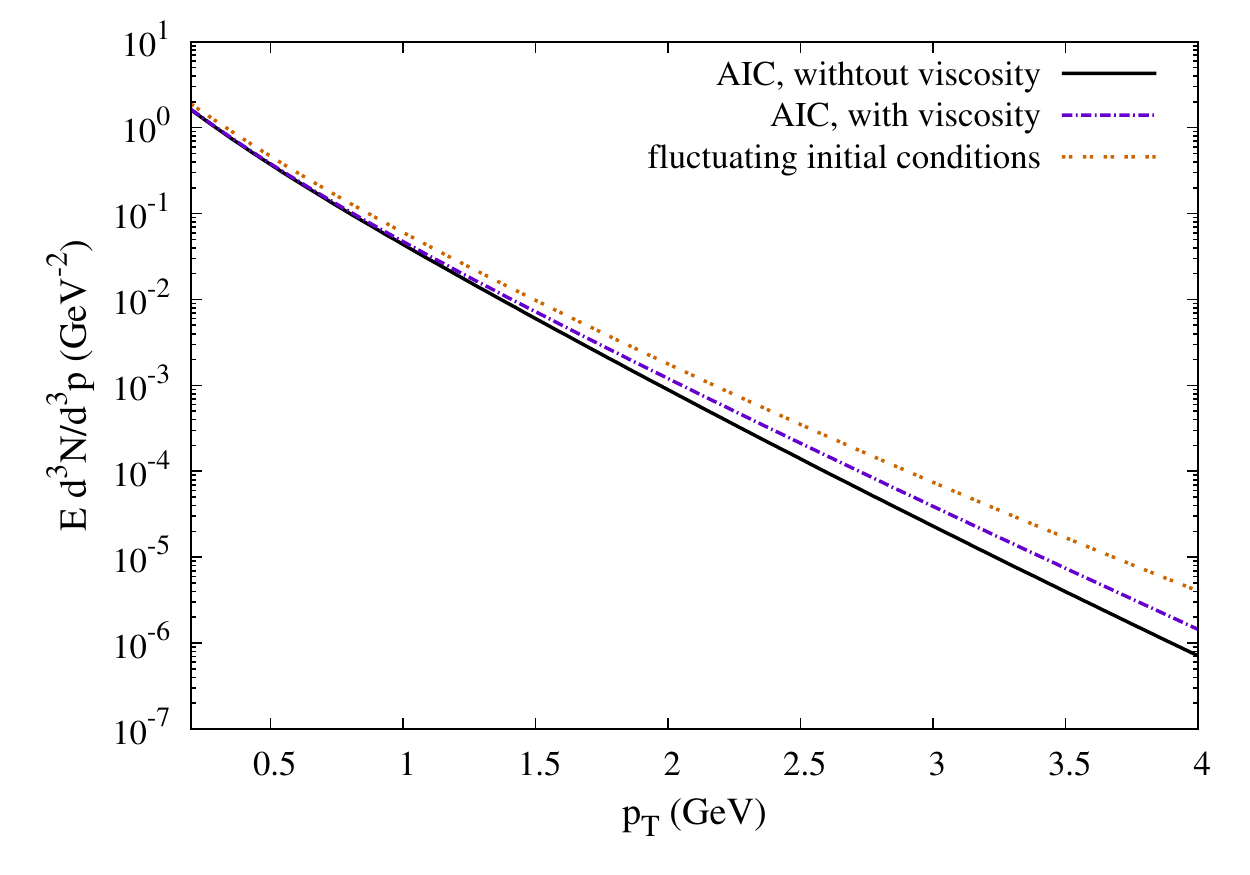}
    \caption{Left panel: Photons from the QGP, using averaged initial conditions for the relativistic hydrodynamics evolution (ideal hydro: solid curve; viscous hydro: dash-dotted curve), and using FICs (with viscosity): double-dotted curve. Right panel: Same as left panel, but but for hadronic gas photons.  This  plot is from Ref. \cite{Dion:2011pp}.}
    \label{FICvisc}
  \end{center}
\end{figure}
\subsection{Fluctuating initial conditions}
Another recent improvement in the understanding of nuclear dynamics,  is that of the role payed by initial conditions and of their effect on finite, odd flow harmonics \cite{Alver:2010gr} for hadrons.  Including Fluctuating Initial Conditions is therefore now an unavoidable feature in the modelling of relativistic nuclear collisions. How will FICs affect the production of electromagnetic radiation? One must first define the nature of the FICs: Here we report on results obtained with a Monte Carlo Glauber approach, previously used for making a successful prediction of the hadronic $v_3$ at RHIC \cite{Schenke:2010rr}. The lumpy initial states germane to FICs lead to harder hadron spectra, owing to a combination of hot spots and of larger blue-shifts from initial gradients  \cite{Qiu:2011iv}. The same arguments will apply to photons, see also Ref. \cite{Chatterjee}. On Figure \ref{FICvisc}, we show the effects of FICs, combined with those of a finite shear viscosity coefficient. An immediate conclusion from that figure: the modification to the spectra from either a finite $\eta$ or from FICs are of similar magnitude. As hydrodynamics is inherently non-linear, it is thus important to have ${\it both}$ ingredients in a single calculation. Again, the effects are largest in regions where pQCD photons will set in. From a theoretical point of view, to make the electromagnetic emissivity consistent with advances in hadronic modelling should however be interpreted as a clear element of progress.

\section{Photon elliptic flow}

\begin{figure}[ht]
  \begin{center}
    \includegraphics[width=7cm]{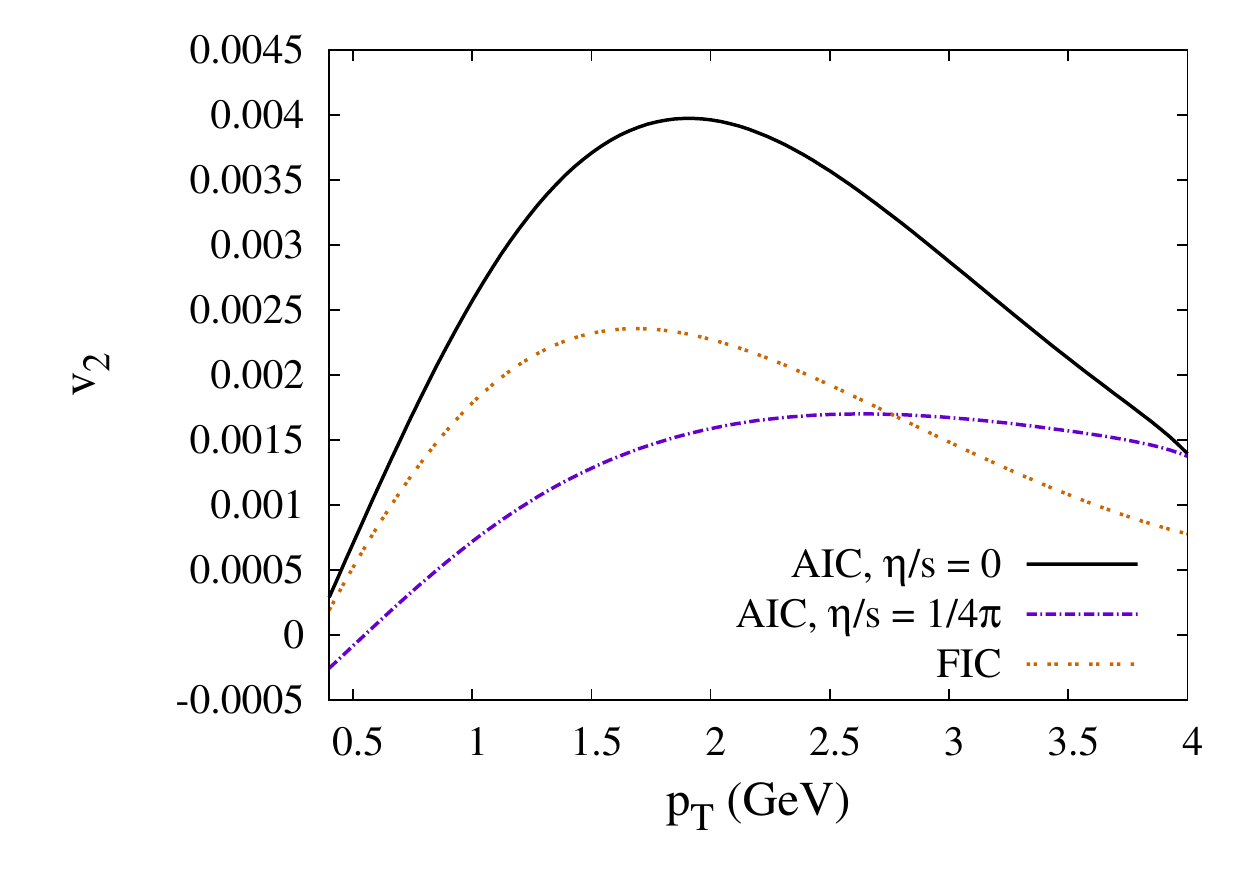}
        \includegraphics[width=7cm]{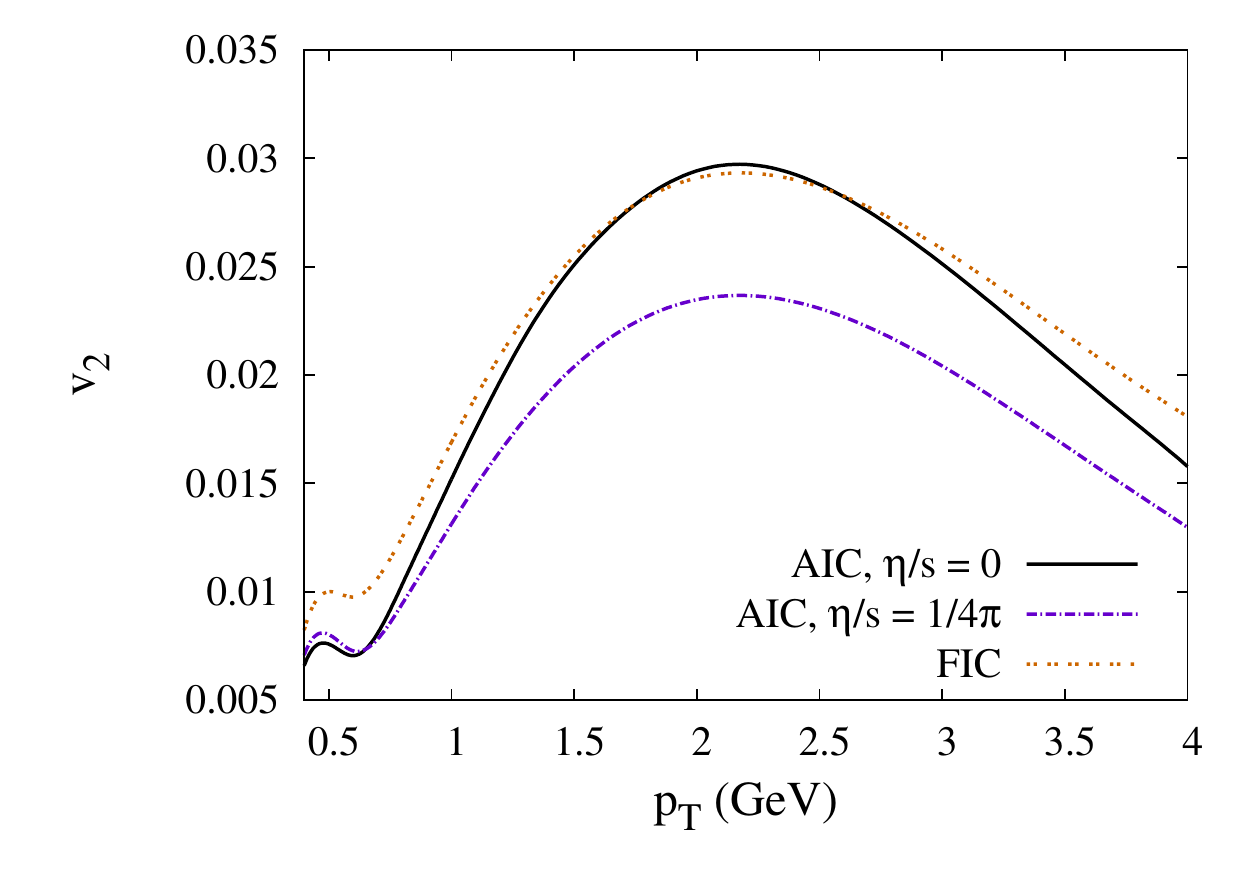}
    \caption{Left panel: Thermal photon $v_2$ from the QGP only. Right panel: Net photon elliptic flow. The solid curve is the result of using inviscid hydrodynamics with averaged initial conditions. The dot-dashed curve shows the effect of a finite shear viscosity coefficient, and the double-dotted curve shows the combined effects of $\eta$ and FICs. This  plot is from Ref. \cite{Dion:2011pp}.}
    \label{v2_FIC}
  \end{center}
\end{figure}

As the measurement of higher flow harmonics have contributed significantly to the success of relativistic hydrodynamics by quantifying the flow anisotropy, it is natural to inquire about the power of similar observables in the electromagnetic sector. As real and virtual photons decouple from the matter once they're produced, and because the photon production rates are mostly $t$-channel dominated (depending on the transverse momentum of the produced photon), their elliptic flow should track the matter momentum anisotropy from early (high-$p_T$ photons) to late (lower-$p_T$ photons) times \cite{Chatterjee:2005de}. Thus, photon elliptic flow is therefore a discriminating observable, complementary to parallel hadronic measurements. A similar story  is predicted to hold true for dileptons \cite{Chatterjee:2007xk}. It is also important to investigate the consequence of out-of-equilibrium effects, as quantified by a finite coefficient of shear viscosity, and of FICs on photon elliptic flow. These effects on $v_2^\gamma$ have been studied individually \cite{Dusling:2009bc,Chaudhuri:2011up}, and together \cite{Dion:2011pp}. They are shown in Figure \ref{v2_FIC}. As for hadrons \cite{Schenke:2010rr}, viscous effects reduce the photon momentum anisotropies: $v_2$ is made smaller by a finite $\eta$. In this centrality class, the FICs make the elliptic flow larger but they also could lead to a reduced $v_2$ in more peripheral events \cite{Schenke:2010rr}. As in the case of spectra, it is therefore important to treat viscous corrections and fluctuating initial states simultaneously, as they compete in magnitude. 

%\begin{figure}[ht]
%  \begin{center}
%    \includegraphics[width=9.5cm]{PHENIX_v2.pdf}
%    \caption{Elliptic in minimum bias collisions of Au + Au at RHIC. The three panels respectively show $v_2$ of $\pi^0$, inclusive and direct  photons, from left to right. The black dots are red squares show two different methods of measuring the reaction plane. This  plot is from Ref. \cite{Adare:2011zr}}
%    \label{PHENIX_v2}
%  \end{center}
%\end{figure}
Preliminary photon $v_2$ data have been available for some years, but were limited to a small range in $p_T$ and to low precision \cite{Adler:2005rg,Sakaguchi:2007zs}. This situation has been firmed up recently, and new results from the PHENIX collaboration have been released \cite{Adare:2011zr}.  At higher $p_T$, the direct photon measurements are consistent with a vanishing elliptic flow, but the experimental uncertainties are currently too large to rule out the small positive contribution from jet fragmentation and/or the small negative contribution from jet-photon conversions \cite{Gale:2009gc}. At low $p_T$ however, the photon elliptic flow is positive, as is that of the strongly interacting source,  but is surprisingly as large as that of the $\pi^0$ flow and increases with centrality. Current models based on relativistic hydrodynamics - which are successful in interpreting hadronic flow - predict much smaller values for photon elliptic flow, as is also clear from Figure \ref{v2_FIC}; see also \cite{Holopainen:2011pd}.  This experimental result has since been confirmed by an independent measurement of the conversion electrons \cite{Drees}. 

\begin{figure}[ht]
  \begin{center}
    \includegraphics[width=7cm]{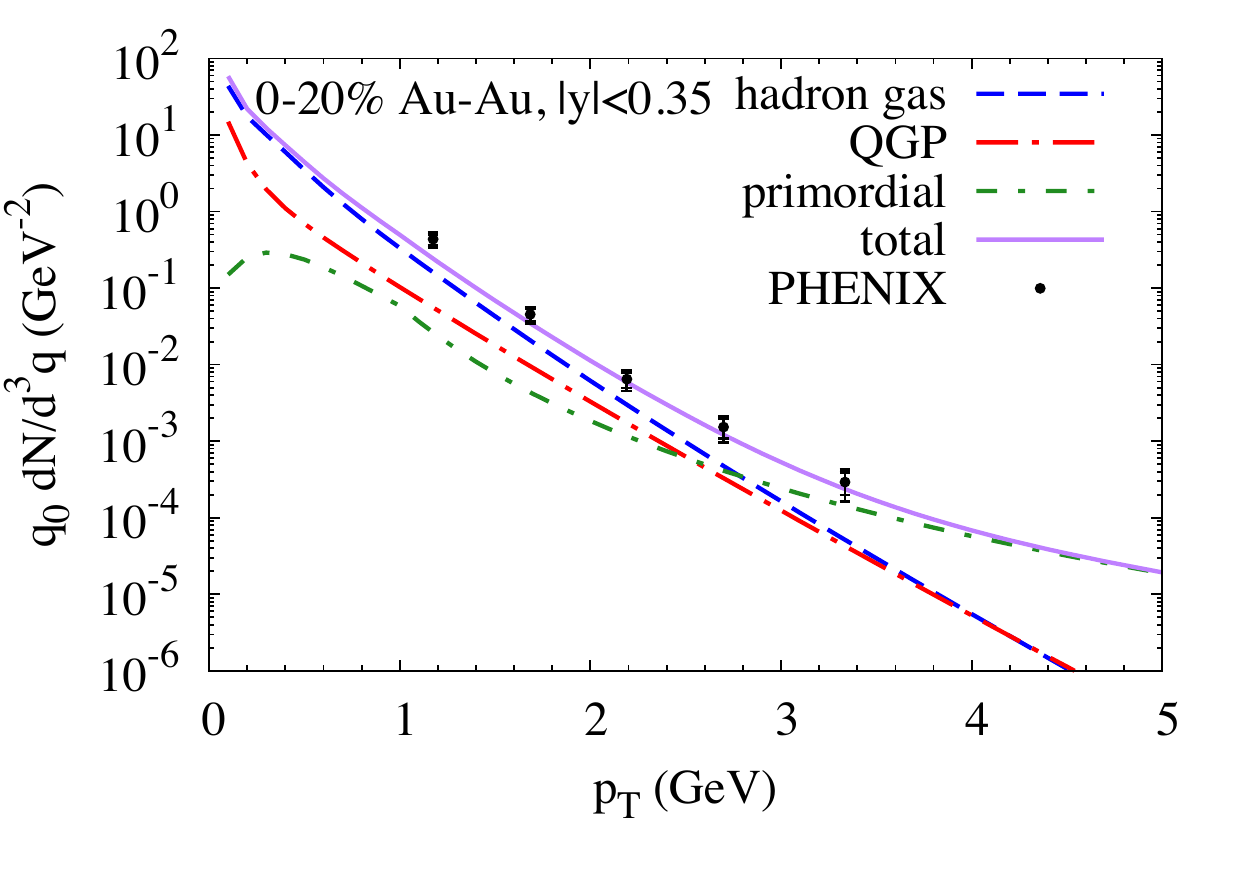}
        \includegraphics[width=7cm]{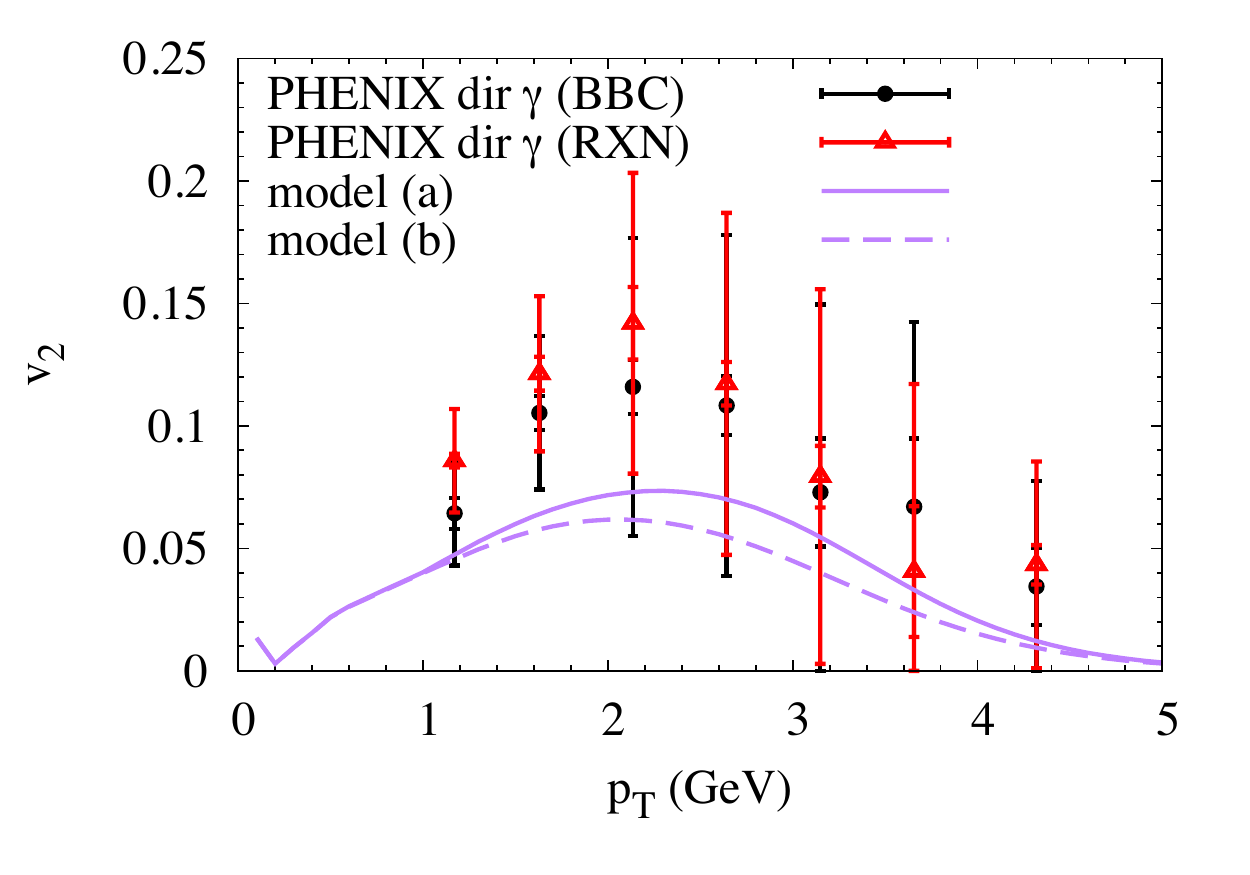}
    \caption{Left panel: Thermal photon yields shown together with PHENIX data. Right panel: the related elliptic flow coefficient. The photons for the partonic phases (QGP) are show, and so are those from the hadronic gas (HG). The other curves are explained in the cited reference.  This  plot is from Ref. \cite{vanHees:2011vb}.}
    \label{fireball}
  \end{center}
\end{figure}

One may start from the $v_2^\gamma$ measurements, and investigate the demands they put on the collisions dynamics. This aspect was pursued in Ref. \cite{vanHees:2011vb}. There, an elliptic fireball expansion constrained by, and consistent with, bulk hadronic data was used. This empirical fireball approach is also consistent with photon measurements at CERN energies performed by the WA98 collaboration, and with dilepton measurements performed by the NA60 collaboration. A configuration which appeared consistent with existing data yields the photon spectrum and elliptic flow shown in Figure \ref{fireball}.  A key ingredient in this work is the need for a large acceleration of the fireball, which blue-shifts the photon spectra to the required hardness. A consequence is then the reduced contribution of the QGP plasma radiation, and the domination of the Hadronic Gas phase. in the HG phase, flow has time to build up and the resulting photon elliptic flow is now large enough to clip the bottom of the experimental error bars. Note however that neither viscous corrections nor fluctuating initial conditions are explicitly considered here, other than through the effective dynamics. Further work is needed to reconcile these results with those obtained through detailed dynamical space-time models. 
\section{Conclusions}
Measurements of electromagnetic radiation from colliding relativistic nuclei have reached an impressive degree of sophistication, even if puzzles remain in what concerns the consistency of different measurements, as well as in their theoretical interpretation. The theory effort has seen considerable consolidation in the modelling of hadronic observables and in the calculation of electromagnetic radiation. More work needs to be done, and this will continue. On the experimental side, next-generation results from RHIC as well as new results on photons and dileptons from the LHC will continue to drive the field forward.

\section*{Acknowledgments} I gratefully acknowledge discussions and collaboration with J. Ghiglieri, U. Heinz, S. Jeon, I. Kozlov, A. Kurkela,  G. D. Moore, J.-F. Paquet, R.  Rapp,  B. Schenke, C.  Shen, D. Teaney, H. van Hees, G. Vujanovic, and C. Young. This work was funded in part by the Natural Sciences and Engineering Research Council of Canada. 



%% References


%\bibliographystyle{elsarticle-num}
\bibliographystyle{model1a-num-names}
\bibliography{Gale_Proceedings_HP2012}
\biboptions{sort&compress}






% \end{thebibliography}

\end{document}